\documentclass[aps,showpacs,preprintnumbers,amsmath, amssymb]{revtex4}

\usepackage{float}
\usepackage{graphics,epsfig}
\usepackage{graphicx}
\usepackage{dcolumn}
\usepackage{bm}

\begin{document}
\baselineskip=0.8 cm

\title{{\bf Analytical investigations on the Maxwell electromagnetic invariant in the spinning and charged horizonless star background}}
\author{Yan Peng$^{1}$\footnote{yanpengphy@163.com}}
\affiliation{\\$^{1}$ School of Mathematical Sciences, Qufu Normal University, Qufu, Shandong 273165, China}

\vspace*{0.2cm}
\begin{abstract}
\baselineskip=0.6 cm
\begin{center}
{\bf Abstract}
\end{center}

We study properties of the Maxwell electromagnetic invariant
in the external region of spinning and charged horizonless stars.
We analytically find that the minimum negative
value of the Maxwell electromagnetic invariant is obtained
on the equator of the star surface.
We are interested in scalar fields
non-minimally coupled to the Maxwell electromagnetic invariant.
The negative enough Maxwell electromagnetic invariant can lead to a
negative effective mass term, which forms a
binding potential well for the scalar field.
It means that the scalar field
coupled to the Maxwell electromagnetic invariant may mostly
exist around the surface of the star on the equator.

\end{abstract}

\maketitle
\newpage
\vspace*{0.2cm}

\section{Introduction}

A characteristic property of black holes is the famous no hair theorem,
which states that static scalar field hairs cannot exist outside
black hole horizons \cite{Bekenstein}-\cite{JBN}. However, this classical theorem can be violated when considering
non-minimally coupled scalar fields.
As an example, static scalar fields non-minimally coupled to
the Maxwell electromagnetic invariant
can exist outside black holes \cite{charge1,charge2,charge3,charge4,charge5,charge6}.
So it is interesting to study the Maxwell
electromagnetic invariant in the curved spacetimes.

Recently, Hod carried out a detailed analytical study on
the physical and mathematical properties
of the Maxwell electromagnetic invariant in the composed Einstein-Maxwell-scalar
field theory outside spinning and charged black holes \cite{SH1}.
Hod found local minimum values of the Maxwell electromagnetic invariant on the horizon and the polar boundary.
Hod further proved that the global minimum value
of the Maxwell electromagnetic invariant is obtained
on the equator of the black hole horizon.
The negative enough Maxwell electromagnetic invariant can lead to a negative effective mass term, which forms a
binding potential well for the scalar field. It means that the non-minimally coupled scalar field
may mostly exist around the horizon of the black hole on the equator.
As the Maxwell electromagnetic invariant plays an important role in
the formation of scalar field hairs, it is of interest to further study properties of
the Maxwell electromagnetic invariant in the background of
spinning and charged horizonless stars.

This paper is to study properties of the Maxwell electromagnetic invariant
outside spinning and charged horizonless stars.
We prove that global minimum values of the Maxwell electromagnetic invariant is obtained
on the equator of the star surface.
This result implies that scalar fields non-minimally coupled to the Maxwell electromagnetic invariant
may mainly exist around the surface of the star on the equator. We will summarize
our main conclusion at the last section.

\section{Investigations on the Maxwell electromagnetic invariant}

We study the Maxwell electromagnetic invariant outside spinning and charged
horizonless stars. The line element of the curved spacetime
is characterized by \cite{KN1,KN2}
\begin{eqnarray}\label{AdSBH}
ds^{2}&=&-\frac{\triangle}{\rho^{2}}(dt-asin^{2}\theta d\phi)^{2}+\frac{\rho^{2}}{\triangle}dr^{2}+\rho^{2}d\theta^{2}+\frac{sin^{2}\theta}{\rho^{2}}[adt-(r^{2}+a^{2})d\phi]^{2}
\end{eqnarray}
with metric functions defined as $\triangle=r^{2}-2Mr+a^{2}+Q^{2}$ and $\rho^{2}=r^{2}+a^{2}cos^{2}\theta$.
The parameters M, J, Q are the ADM mass, angular momentum and electric charge of the spacetime respectively.
And we label the star surface with the radial coordinate $r=r_{s}$.

The coupling between scalar fields and the Maxwell electromagnetic invariant
plays an important role in the existence of scalar field hairs in curved spacetimes.
The composed Einstein-Maxwell-scalar field gravity is described by the action \cite{charge1,charge2,charge3,charge4,charge5,charge6}
\begin{eqnarray}\label{AdSBH}
S=\int d^{4}x\sqrt{-g}[R-2\nabla_{\alpha}\nabla^{\alpha}\psi-2\mu^{2}\psi^{2}-f(\psi)F],
\end{eqnarray}
where $f(\psi)$ is the function $f(\psi)=1-\alpha \psi^{2}$ with $\alpha$  describing the strength of coupling
between the scalar field and the Maxwell electromagnetic invariant.
$\mu^{2}$ is the square mass of the scalar field.
$R$ is the scalar curvature and $F$ is
the Maxwell electromagnetic invariant expressed by the
electromagnetic tensor $F_{\mu \nu}$ in the form \cite{EM1,EM2}
\begin{eqnarray}\label{AdSBH}
F(r,\theta,a,Q)=F_{\mu\nu}F^{\mu\nu}=-\frac{2Q^{2}(r^{4}-6r^{2}a^{2}cos^{2}\theta+a^{4}cos^{4}\theta)}{(r^{2}+a^{2}cos^{2}\theta)^{4}}
\end{eqnarray}
with the radial coordinate $r\in [r_{s},\infty)$ and the polar angular coordinate
$\theta\in [0,\pi]$.

The action (2) yields the scalar field differential equation
\begin{eqnarray}\label{BHg}
\nabla^{\nu}\nabla_{\nu}\psi=\mu^{2}_{eff}\psi,
\end{eqnarray}
where $\mu^{2}_{eff}$ is the effective mass $\mu^{2}_{eff}=\mu^{2}-\frac{1}{2}\alpha\cdot F$.
For the physical parameter $\alpha<0$, the effective mass becomes negative
for negative enough Maxwell electromagnetic invariant.
The presence of a negative effective mass term introduces
a potential well for the scalar field,
leading to the formation of scalar field hairs.
So the scalar field may exist mainly around minimum points
of the negative Maxwell electromagnetic invariant.

Then we try to search for the mostly negative Maxwell electromagnetic invariant in the region
\begin{eqnarray}\label{BHg}
r\in [r_{s},\infty),~\theta\in[0,\pi].
\end{eqnarray}

We will prove that the mostly negative
value of the Maxwell electromagnetic invariant must be on the equator.
Defining a new parameter $y=a^{2} cos^{2}\theta$, the formula (3) can be expressed as
\begin{eqnarray}\label{AdSBH}
F(r,\theta,a,Q)=-\frac{2Q^{2}(r^{4}-6r^{2}y+y^{2})}{(r^{2}+y)^{4}}.
\end{eqnarray}

For the fixed radial coordinate $r\geqslant r_{s}$, the mostly negative value of the Maxwell electromagnetic invariant
may be obtained on extremum points or the boundary of the polar angular coordinate.
From $\frac{\partial F}{\partial y}=0$, one finds extremum points as: $y=(5-2\sqrt{5})r$ and $y=(5+2\sqrt{5})r$.
At extremum points and boundaries $y=0$ and $y=\infty$,
the values of the Maxwell electromagnetic invariant are
\begin{eqnarray}\label{AdSBH}
F(r,y=0)=-\frac{2Q^{2}}{r^{4}},
\end{eqnarray}
\begin{eqnarray}\label{AdSBH}
F(r,y=(5-2\sqrt{5})r)=-\frac{Q^{2}(\sqrt{5}-2)}{(\sqrt{5}-3)^{4}r^{4}}\approx \frac{0.6931 Q^{2}}{r^{4}},
\end{eqnarray}
\begin{eqnarray}\label{AdSBH}
F(r,y=(5+2\sqrt{5})r)=-\frac{Q^{2}(\sqrt{5}+2)}{(\sqrt{5}+3)^{4}r^{4}}\approx -\frac{0.0056 Q^{2}}{r^{4}},
\end{eqnarray}
\begin{eqnarray}\label{AdSBH}
F(r,y\rightarrow \infty)=0.
\end{eqnarray}

Along a fixed radial coordinate r, the mostly negative Maxwell electromagnetic invariant is obtained at
$y=0$ (or $cos^{2}\theta=0$) as
\begin{eqnarray}\label{AdSBH}
F(r,cos^{2}\theta=0)=-\frac{2Q^{2}}{r^{4}}.
\end{eqnarray}

In other words, the Maxwell electromagnetic invariant is more negative on the equator as
\begin{eqnarray}\label{AdSBH}
F(r,cos^{2}\theta\neq0)>F(r,cos^{2}\theta=0).
\end{eqnarray}

According to relation (11), the Maxwell electromagnetic invariant
on the equator is more negative for smaller radial coordinate r.
For fixed radial coordinate $r>r_{s}$, we obtain the relation
\begin{eqnarray}\label{AdSBH}
F(r,cos^{2}\theta=0)>F(r_{s},cos^{2}\theta=0)=-\frac{2Q^{2}}{r_{s}^{4}}.
\end{eqnarray}

Relations (12) and (13) yield the following inequality
\begin{eqnarray}\label{AdSBH}
F(r,cos^{2}\theta)>F(r_{s},cos^{2}\theta=0)
\end{eqnarray}
for a general coordinate ($r$,$cos^{2}\theta$)$\neq$($r_{s}$,$cos^{2}\theta=0$).

So the mostly negative value of the Maxwell
electromagnetic invariant is around the surface of the star
on the equator ($r=r_{s}$ and $\theta=\frac{\pi}{2}$).
The negative enough Maxwell electromagnetic invariant
can introduce negative effective mass and form a
binding potential well for the scalar field,
leading to the existence of scalar field hairs
around the surface of the star on the equator.
Now we try to obtain the condition for the onset of the spontaneous scalarization phenomenon.
From the scalar field equation (4) and the inequalities (13), the scalarization condition
is determined by the relation $\mu^{2}_{eff}\leqslant0$, which can be expressed as
\begin{eqnarray}\label{AdSBH}
\mu^{2}-\frac{1}{2}\alpha(-\frac{2Q^2}{r_{s}^{4}})\leqslant0.
\end{eqnarray}

According to (15), one concludes that scalar fields with mass
\begin{eqnarray}\label{AdSBH}
\mu^{2}\leqslant-\frac{\alpha Q^2}{r_{s}^{4}}
\end{eqnarray}
can exist in the external region of spinning and charged stars.

\section{Conclusions}

We investigated on the properties of the Maxwell electromagnetic invariant
in the external region of spinning and charged horizonless stars.
We are interested in the gravity system with scalar fields
non-minimally coupled to the Maxwell electromagnetic invariant.
The Maxwell electromagnetic invariant plays an important role
in the existence of scalar field hairs.
According to the scalar field equation,
a negative enough Maxwell electromagnetic invariant can lead to
a negative effective mass term.
And the negative effective mass term forms a
potential well to bind the scalar field.
So the scalar field is more easier to exist where
the Maxwell electromagnetic invariant is more negative.
We analytically found that the global minimum negative
value of the Maxwell electromagnetic invariant is obtained
on the equator of the star surface,
which means that the scalar field
coupled to the Maxwell electromagnetic invariant may mostly
exist around the surface of the star on the equator.

\begin{acknowledgments}

This work was supported by the Shandong Provincial Natural Science Foundation of China under Grant
No. ZR2022MA074. This work was also supported by a grant from Qufu Normal University
of China under Grant No. xkjjc201906, the Youth Innovations and Talents Project of Shandong
Provincial Colleges and Universities (Grant no. 201909118), Taishan Scholar Project of Shandong Province (Grant No.tsqn202103062)
and the Higher Educational Youth Innovation Science
and Technology Program Shandong Province (Grant No. 2020KJJ004).

\end{acknowledgments}


\begin{thebibliography}{99}







\bibitem{Bekenstein}
J. D. Bekenstein, Transcendence of the law of baryon-number conservation in black hole physics, Phys. Rev. Lett. 28, 452 (1972).




\bibitem{Chase}
J. E. Chase, Event horizons in Static Scalar-Vacuum Space-Times, Commun. Math. Phys. 19, 276 (1970).



\bibitem{C. Teitelboim}
C. Teitelboim, Nonmeasurability of the baryon number of a black-hole, Lett. Nuovo Cimento 3, 326 (1972).





\bibitem{Ruffini-1}
R. Ruffini and J. A. Wheeler, Introducing the black hole, Phys. Today 24, 30 (1971).




\bibitem{W.K.H}
W.K.H. Panofsky, Needs Versus Means In High-energy Physics, Phys. Today 33(1980)24-33.




\bibitem{MH1}
M. Heusler, A No hair theorem for selfgravitating nonlinear sigma models, J. Math. Phys. 33(1992)3497-3502.



\bibitem{MH2}
Markus Heusler, A Mass bound for spherically symmetric black hole space-times, Class. Quant. Grav. 12(1995)779-790.


\bibitem{JBN}
J.D. Bekenstein, Novel ``no-scalar-hair'' theorem for black holes, Phys. Rev. D 51(1995)no.12,R6608.






\bibitem{charge1}
Carlos A.R. Herdeiro, Eugen Radu, Nicolas Sanchis-Gual, Jos\'{e} A. Font,
Spontaneous Scalarization of Charged Black Holes, Phys. Rev. Lett. 121(2018)101102.




\bibitem{charge2}
Pedro G.S. Fernandes, Carlos A.R. Herdeiro, Alexandre M. Pombo, Eugen Radu, Nicolas Sanchis-Gual,
Spontaneous Scalarisation of Charged Black Holes: Coupling Dependence and Dynamical Features,
Class. Quant. Grav. 36(2019)no.13,134002.



\bibitem{charge3}
S. Hod, Spontaneous scalarization of charged Reissner-Nordstr\"{o}m black holes:
Analytic treatment along the existence line, Physics Letters B 798(2019)135025.




\bibitem{charge4}
Yun Soo Myung, De-Cheng Zou,
Instability of Reissner-Nordstr\"{o}m black hole in Einstein-Maxwell-scalar theory,
Eur. Phys. J. C 79(2019)no.3,273.



\bibitem{charge5}
Yun Soo Myung, De-Cheng Zou,
Stability of scalarized charged black holes in the Einstein-Maxwell-Scalar theory,
Eur. Phys. J. C 79(2019)no.8,641.



\bibitem{charge6}
De-Cheng Zou, Yun Soo Myung,
Scalar hairy black holes in Einstein-Maxwell-conformally coupled scalar theory,
arXiv:1911.08062[gr-qc].





\bibitem{SH1} Shahar Hod, Analytic study of the Maxwell electromagnetic invariant in spinning and charged Kerr-Newman black-hole spacetimes,
JHEP09(2023)140.






\bibitem{KN1} C. W. Misner, K. S. Thorne, and J. A. Wheeler, Gravitation,(W. H. Freeman, San Francisco,
1973).



\bibitem{KN2} S. Chandrasekhar, The Mathematical Theory of Black Holes,(Oxford University Press, New
York, 1983).








\bibitem{EM1}  T. Adamo and E.T. Newman, The Kerr-Newman metric: A Review, arXiv:1410.6626 .




\bibitem{EM2}  I. Dymnikova and E. Galaktionov, Advan. in Math. Phys. 2017,1035381 (2017)
[https://www.hindawi.com/journals/amp/2017/1035381/]; I. Dymnikova and E. Galaktionov,
Universe 5,205 (2019).












\end{thebibliography}
\end{document}